\documentclass[twocolumn,twoside,slac_two]{revtex4wtf}
\usepackage{graphicx}
\usepackage{fancyhdr}
\usepackage{booktabs} 

\begin{document}

\title{Missing Gamma-Rays from kpc-scale AGN Jets: A Test of the IC/CMB Model}

%

\author{Eileen T. Meyer}
\affiliation{Space Telescope Science Institute, Baltimore, MD 21218, USA}
\affiliation{University of Maryland Baltimore County, Baltimore, MD 21250, USA}
\author{Markos Georganopoulos}
\affiliation{University of Maryland Baltimore County, Baltimore, MD 21250, USA}

\author{W. B. Sparks}
\affiliation{Space Telescope Science Institute, Baltimore, MD 21218, USA}

\author{Leith Godfrey}
\affiliation{ASTRON Netherlands Institute for Radio Astronomy, The Netherlands}

\author{Eric Perlman}
\affiliation{Florida Institute of Technology, Melbourne, FL 32901, USA}




\begin{abstract}
The physical origin of the X-ray emission in powerful quasar jets has
been a long-standing mystery. Though these jets start out on the
sub-pc scale as highly relativistic flows, we do not have any direct
measurement of their speeds on the kpc scale, where the vast distances
from the core necessitate in situ particle acceleration. If the jets
remain highly relativistic on kpc scales, then the X-rays could be due
to inverse-Compton upscattering of CMB photons.  However, the IC/CMB
explanation predicts a high level of gamma-ray emission, which should
be detectible by the \emph{Fermi}/LAT. We have searched for and ruled
out this emission at a high level of significance for the well-known
sources 3C 273 and PKS 0637-752, suggesting the X-rays are
synchrotron, though of unknown origin. These recent results with Fermi
also suggest that the kpc-scale jets in powerful quasars are
significantly slower than have been presumed under the IC/CMB
model. I will discuss the surprising implications of these findings
for the energetics and radiative output of powerful quasars as well as
their impact on their environment.
\end{abstract}

\maketitle

\thispagestyle{fancy}

\begin{figure}[!ht]
\begin{center}
\includegraphics[width=3in]{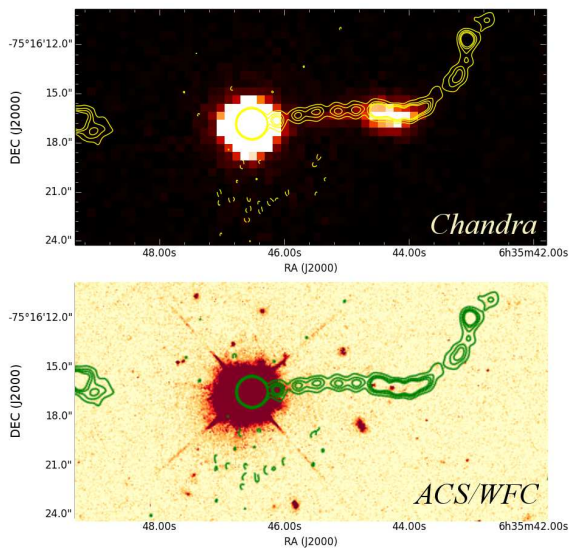}
\caption{\emph{Upper Panel:} \emph{Chandra} X-ray image of
  PKS~0637-752, with ATCA 17 GHz radio contours overlaid. \emph{Lower
    Panel:} Optical image of PKS~0637-752, taken with ACS/WFC on HST
  (F475W filter) with ATCA 17 GHz radio contours overlaid.}
\end{center}
\label{fig:pks0637_panels}
\end{figure}

In August 1999, the \emph{Chandra} X-ray Observatory observed its
first celestial target, quasar PKS~0637-752, during the initial
focusing of the telescope \citep{schwartz2000,chartas2000}. Along with
the bright quasar core, \emph{Chandra} unexpectedly detected X-rays
from the kilo-parsec scale relativistic jet (previously known from
radio imaging, Figure~\ref{fig:pks0637_panels}). Unlike the
synchrotron spectrum of lower-power FR I jets like M87 which easily
extend up to X-ray energies \citep[e.g.][]{wilson2002}, the
synchrotron spectrum of powerful quasar jets (including PKS~0637-752)
generally peak at or below the IR/Optical band. The X-rays detected in
the kpc-scale jet of PKS~0637-752 were orders of magnitude brighter
than expected from the radio-optical synchrotron spectrum, or indeed
from either synchrotron self-Compton (SSC) or inverse Compton
upscattering of ambient CMB photons (IC/CMB) under equipartition
conditions \citep{chartas2000}.  Further, the X-ray spectrum of the
jet was remarkably hard, with a photon index of $1.76\pm0.1$.

Proper motions measurements of sub-parsec scale jets of powerful
quasars with Very Long Baseline Interferometry (VLBI) have detected
superluminal proper motions which imply that these jets start out
highly relativistic, with Lorentz factors ($\Gamma$) of 10-50
\citep{jor05,lis09}. Though it had long been supposed based on
population studies that jets decelerate and are at most mildly
relativistic by the time they reach the kpc scale
\citep[e.g.][]{arshakian2004,mullin2009}, no direct measurements have
confirmed this. \cite{tav00} and \cite{cel01} thus suggested
that the X-rays from the jet in PKS~0637-752 could be explained by
IC/CMB emission if the jet \emph{remained} highly relativistic
($\Gamma\sim$10), and was pointed at a fairly small angle to our line
of sight (6$^\circ$). This produces a much larger Doppler boosting
factor ($\delta\sim$10) and enables the IC/CMB X-rays to match the
observations.

Over the past decade and a half since the launch of \emph{Chandra},
dozens more kpc-scale quasar jets with anomalously hard and/or high
X-rays have been detected
\citep[e.g.][]{sambruna2001_3c273,sambruna2002,aneta2003,sambruna2004,marshall2005,harris2006,siemiginowska2007,marshall2011,kharb2012,godfrey2012_2101}. The
IC/CMB model has been by far the most popular explanation of these
X-rays, though problems have been noted \citep{har06}.  Besides the
unconfirmed fast speeds required on the kpc scale, IC/CMB often
requires the jet to be pointed very close along our line-of-sight,
leading to a deprojected jet length longer than 1 Mpc, the upper limit
for jets observed in the plane of the sky. Further, the electrons
responsible for upscattering the CMB into the \emph{Chandra} band are
at much lower energies than are traced by radio observations. This
extension of the electron energy distribution is energetically costly,
in some cases leading to `super-Eddington' jet power requirements
\citep{der04,uch06}. These problems lead to the suggestion that the
X-rays could alternatively be synchrotron emission from a second
electron population in the jet, albeit of unknown origin
\citep{harris2004,kataoka2005,har06,jes06,uch06}. he fundamental
problem up to now has been that fitting the radio-to-X-ray spectral
energy distribution (SED) alone cannot distinguish between the IC/CMB
and synchrotron explanations for the X-rays \citep{cara2013}. The
difference in power requirements between the two mechanisms is great,
as is the extremely different idea of jet structure that they
imply. Discriminating between these models is essential to make
progress on the actual impact of jets on their environment.

\citet{geo06}, hereafter \textbf{G06}, suggested that \emph{Fermi}
Large Area Telescope (LAT) observations could confirm or rule out the
IC/CMB mechanism for the X-rays, by detecting (or not) the high level
of gamma-ray emission this mechanism requires. We have previously looked for this
gamma-ray emission from the jet of 3C~273, and ruled out IC/CMB
gamma-rays from (the brightest) knot A alone at the $>$95\% level, and
from knots A through D1 combined at the $>$99.9\% level
\citep[][hereafter \textbf{M14}]{mey14}.

At this meeting, we reported new \emph{Fermi} observations of PKS~0637-752
which show that the expected steady gamma-ray emission from the IC/CMB
mechanism is also ruled out by deep upper limits at the 99.9\% level. We
also present updated limits for 3C ~273, showing that the expected
gamma-rays from IC/CMB are now ruled out at the 99.99\% level in more
than one \emph{Fermi} energy band. We will briefly discuss the
implications of these measurements on two fronts. First, we show that
a second synchrotron component is the only likely scenario left to
explain the X-rays in these jets. Secondly, we find that irrespective
of the X-ray emission mechanism, the deep upper limits at GeV energies
place interesting constraints on the Doppler beaming factors which
implies that these jets are not highly relativistic on the kpc scale.
We will discuss the surprising implications of slow, synchrotron X-ray
jets on our understanding of the total radiative output of quasars,
especially at TeV energies.

\section{methods}
\label{sec:methods}

\subsection{The \emph{Fermi} Test of IC/CMB}

As first noted by G06, the shape of the IC/CMB spectrum is constrained
to match the synchrotron spectrum, with a shift in frequency and
luminosity solely determined by the factor $B/\delta$ where $\delta$
is the Doppler beaming factor and $B$ the magnetic field
strength. From G06, we have:

\begin{equation}
\frac{\nu_c}{\nu_s} = \frac{\nu_\mathrm{CMB}\delta^2\gamma^2}{eB\delta\gamma^2/[2\pi m_e c(1+z)]}
\end{equation}

\begin{equation}
\frac{L_c}{L_s} = \frac{32\pi U_\mathrm{CMB}(1+z)^4\delta^4}{3(B\delta)^2}, 
\end{equation}
where $\nu_c$ and $\nu_s$ ($L_c$, $L_s$) are the observed Compton and
synchrotron frequencies (luminosities) emitted by electrons of Lorentz
factor $\gamma$, $e$ and $m_e$ are the electron charge and mass, and
$\nu_\mathrm{CMB}$ = 1.6$\times 10^{11}$ Hz is the CMB peak frequency
at z = 0. However, if the observed X-ray fluxes are to be produced by
the IC/CMB mechanism, then the value of $B/\delta$ is already uniquely
determined by the requirement to match the X-ray flux level, at which
point there is no freedom at all in the rest of the spectrum.  The
peak of the IC/CMB spectrum will fall in the GeV band. Note that this prediction is not
predicated on any particular (e.g., equipartition) magnetic field
strength.

The \emph{Fermi}/LAT lacks the spatial resolution to detect the jet
separately from the gamma-ray bright quasar core, which is only 10$''$
away -- the \emph{Fermi}/LAT 68\% containment radius is on the order
of tenths of a degree to degrees. However, in powerful quasars the
inverse Compton core emission generally peaks at a few MeV, producing
a soft, and extremely variable spectrum in the \emph{Fermi} band,
with long periods of relative quiescence. Indeed,
PKS~0637-752 was detected in the 2nd \emph{Fermi} source catalog
\citep[2FGL,][]{nolan2012} as both a very soft (photon index of
$\Gamma_p$=2.71) and highly variabile source (variability index =
347). In contrast, the IC/CMB emission from the large scale jet is
expected to be harder and completely non-variable. The latter property
allows us to combine the \emph{Fermi} data taken only when the quasar
core is in a low state to try to detect or place limits on the IC/CMB
emission.

\subsection{Fermi Analysis of PKS 0637-752}

We first combined the all-sky weekly LAT event and spacecraft files
for weeks 9 through 325 of the \emph{Fermi} mission, corresponding to
Fermi Mission Elapsed Time (MET) from 239557417 to 430608212 and
calendar dates 4 August 2008 to 24 August 2014. In order to analyze
the region around PKS~0637-752, we used the publicly available
`quickAnalysis' script. The public scripts mentioned here are
available at http://fermi.gsfc.nasa.gov/ssc/data/analysis/user/. to
run the \emph{Fermi} analysis tools and generate the filtered event
file, livetime cube, and exposure map, using a region of interest
(ROI) of 10$^\circ$ and an otherwise default configuration. The
starting source list was generated from the publicly available
make2FGLxml script, which generates the xml file pre-populated with
2FGL catalog sources. We used a binned maximum Likelihood to get an
initial fit for all the catalog sources in our ROI. We also included
sources a further 5$^\circ$ out from our ROI, but always fixed to the
catalog values. PKS~0637-752 was detected with a very high
test-statistic (TS, roughly significance squared) value of 289, a
100~MeV to 100~GeV photon flux of 3.16$\times 10^{-8}$ s$^{-1}$
cm$^{-2}$, and a photon index $\Gamma_p$ = 2.64, similar to the value
reported in the 2FGL catalog.

\begin{figure}[!b]
\begin{center}
\includegraphics[width=3.2in]{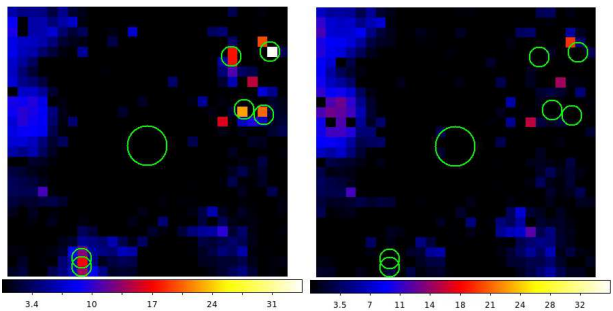}
\caption{
\label{fig:initTSmap}\emph{Left: }An initial TS map of the region around
  PKS~0637-752, showing the excess TS present in 0.5$^\circ$ pixels
  over the best-fit likelihood model using 2FGL catalog sources.  The
  large circle marks the position of PKS~0637$-$752. The smaller
  circles mark the positions of the six new sources not present in the
  2FGL catalog (corresponding to regions of excess TS with pixel
  values $>$20). \emph{Right: } The updated TS map (same FOV and
  binning) after the six sources were localized and fit with a binned
  likelihood.}
\end{center}
\end{figure}

\begin{table}
\caption{\label{table:newsources} New sources in ROI of Targets}
\centering
\begin{tabular}{llc|llc}
\toprule
\underline{0637-752} & & & \underline{3C 273} & & \\
RA$_{2k}$ & Dec$_{2k}$ & TS & RA  & Dec & TS  \\
(deg) & (deg) &  & (deg) & (deg) & \\
\midrule
  \phantom{0}82.43046 & -72.74572 &           57.5 & 192.82676 &           -2.00263 & \phantom{0}70.7 \\
  \phantom{0}81.13040 & -69.60575 &           60.5 & 190.96827 &           -2.29561 & \phantom{0}82.3 \\
  \phantom{0}78.98804 & -72.72570 &           28.6 & 187.15935 &           -3.29476 & \phantom{0}60.2 \\
  \phantom{0}86.33797 & -70.34640 &           39.6 & 184.47577 &           -0.48239 & \phantom{0}92.5 \\
            119.70430 & -80.70430 & \phantom{0}9.9 & 193.43690 & \phantom{-}3.47391 &           224.6 \\
            118.83300 & -80.32970 &           16.1 & 192.63313 & \phantom{-}2.25116 & \phantom{0}65.7 \\
\bottomrule
\end{tabular}
\end{table}

We checked for additional significant sources within 7$^\circ$ of
PKS~0637-752 but not in the 2-year LAT catalog by making a TS residual
map.  Rough starting positions of apparent new significant sources
were measured from the TS map by hand, only considering as candidates
those with a central pixel value (TS) $>$ 20. Each new source was
added to the XML model file as a powerlaw, and an initial spectral
parameter fit derived via binned likelihood. We then refined the
position of each source one at a time while the spectral parameters
were held fixed.  The existing tool, {\tt gtfindsrc}, only works for
unbinned likelihood analysis, so we built our own binned version of
the tool which works in the same way.  Using the frozen model, we used
the python {\tt minimize} function in the {\tt scipy} package
(L-BFGS-B method) to optimize the log-likelihood value versus the RA
and Dec position, given a reasonable range of about 1 degree around
the starting positions noted by hand. For the 7$^\circ$ ROI around
PKS~0637-752, six new sources were added to the model, and an updated
TS map run from this larger source list shows that the excess TS
previously seen is now gone (right panel of Figure~2). A list of the
new sources with their location and TS value is given at left in
Table~\ref{table:newsources}.

\begin{figure}[!b]
\begin{center}
\includegraphics[width=3.2in]{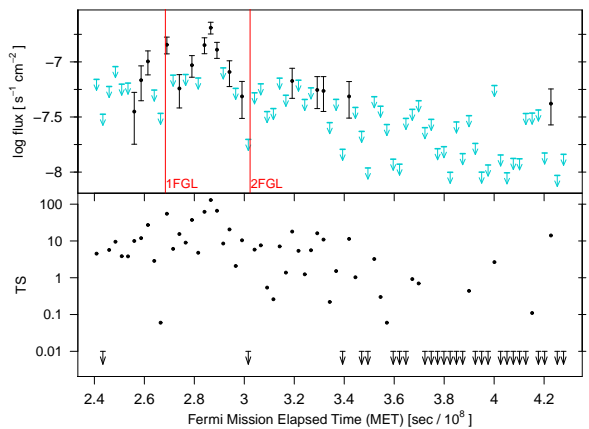}
\caption{\label{fig:0637_lightcurve}\emph{Upper: }Lightcurve of
  PKS~0637-752. The total 100 MeV - 100 GeV photon flux for
  PKS~0637-752 in 10.5-day (total GTI) bins versus the mean Mission
  Elapsed Time (MET) of the bin. Upper limits are shown where
  TS$<$10. The red verticle lines show the end-time for the 1FGL and
  2FGL catalogs. \emph{Lower: } The TS value corresponding to the same
  bin as above. }
\end{center}

\end{figure}
PKS~0637-752 is a significant \emph{Fermi} source, and was detected in
the 2nd \emph{Fermi} catalog, as 2FGL~J0635.5-7516. Our approach to
detecting and/or setting limits on the IC/CMB gamma-ray emission
exploits the variability of the blazar core which cannot be spatially
resolved separately from the large-scale jet due to the poor angular
resolution of \emph{Fermi}. During times when the blazar is quiescent,
the hard, steady emission from IC/CMB will either appear as a steady
plateau, or else the upper limits generated will place constraints on
the level of the IC/CMB emission.  In order to build a lightcurve of
the core, the full 6-year dataset is divided into bins of equal good
time interval (GTI) time, totalling 10.5 days.  We then used our
updated (2FGL + 6 new) model described above and ran a binned
likelihood to fit PKS~0637-752 as a power-law source, with sources
more than 5$^\circ$ away fixed.  The resulting lightcurve for the core
over the full time range is shown in Figure~3, with the 100 MeV - 100
GeV photon flux shown on top and the corresponding TS shown below. The
clear variability in the light curve indicates that the total
\emph{Fermi} flux is dominated by the core.

\begin{figure}[!b]
\begin{center}
\includegraphics[width=3.2in]{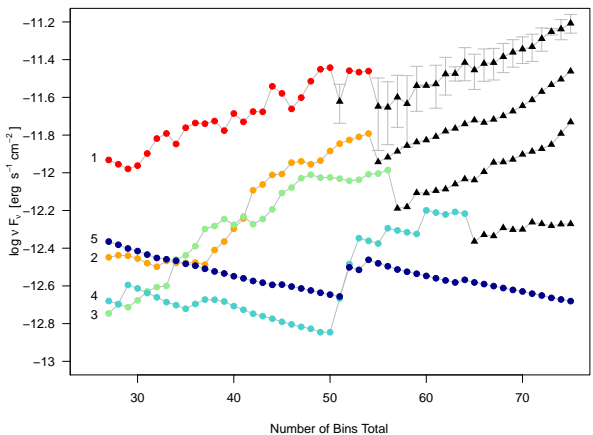}
\caption{\label{fig:pks0637_limits}The results of the progressive binning
  analysis on PKS~0637-752. The upper axis gives the $\nu F_\nu$ flux while lower axis
  gives the total number of bins combined (where ordering is based on
  TS value and not time order), starting from the 27 bins with
  TS$<$0.01. The points give the upper limits (colored dots) or
  detected fluxes (TS$>$10, black triangles) for each of the five
  \emph{Fermi} energy bands (red, orange, green, cyan, navy from
  lowest to highest energy). Errors on the detected fluxes are only
  shown for the lowest-energy bin for clarity but are similar across
  bins. For the highest-energy bin (10 - 100 GeV) no detection is ever
  made. The increasing fluxes indicate that the quasar core is being
  detected in the other bands.}
\end{center}
\end{figure}

\begin{figure*}[t]
\begin{center} 
\includegraphics[width=6in]{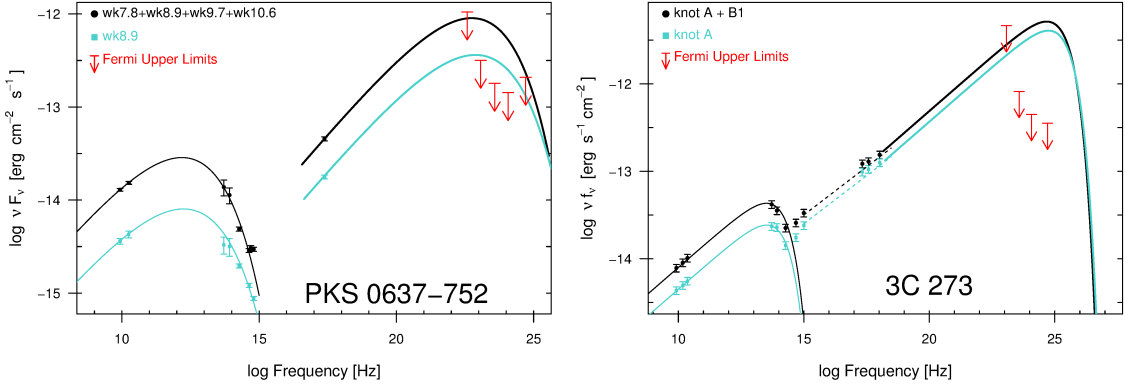}
\caption{\label{fig:both_results}\emph{Left: }The SED for the
  large-scale jet of PKS~0637-752.  Data for the four brightest X-ray
  detected knots combined is shown as black points. The SED of the
  X-ray brightest knot, wk8.9, is plotted with blue points.
  \emph{Right: }The SED for the knots of 3C~273, with black points for
  knots A and B1 combined, and blue for X-ray brightest knot A only. For both plots, the
  \emph{Fermi} 95\% upper limits are shown in red.
}
\end{center}
\end{figure*}

We next began a `progressive binning' analysis, in which the
lightcurve bins were ordered from lowest to highest TS value. Of the
entire set of 75 time bins, 27 showed a TS level consistent with zero
for the location of PKS 0637-752 (upper limits in lower panel of
Figure~\ref{fig:0637_lightcurve}).  Starting from these 27 bins
combined, we progressively combine the event files for the lowest bins
plus the next lowest bin in TS, at each step optimizing the fit of
PKS~0637-752 and the sources within 5$^\circ$ with a binned
likelihood. We repeated adding the next-highest bin and getting the
maximum likelihood fit until all bins had been added together. Note
that this re-combining of the lightcurve in a discontinuous way is
appropriate for deriving a limit on the large-scale jet because the
IC/CMB emission is predicted to be completely non-variable, and thus
there is no risk of any selection effect via variability. The variable
core clearly clearly dominates the flux levels determining the
ordering, and is disconnected from the jet in any case.  At each step
we evaluated the TS and flux level in the five canonical \emph{Fermi}
energy bands of 0.1-0.3, 0.3-1, 1-3, 3-10 and 10-100 GeV, calculating
the 95\% upper limit flux value when TS$<$10 in any given
bin. Previous work on 3C~273 has shown that the exact ordering of the
bins (whether by using the TS value or the total flux or upper limit
value for the bin) does not significantly affect the resulting upper
limits (M14).

As shown in Figure~\ref{fig:pks0637_limits}, the highest energy bands
gave upper limits which decreased with the increasing exposure as more
bins are added, where we have color-coded the upper limits in the 5
energy bands, and black triangles indicate a significant detection in
the band. The decrease in upper limits sometimes going faster than
1/$\sqrt{t}$ and `jumpy' behavior is expected in the case of very low
backgrounds and low count rates. Note that in the lower-energy bins,
where the PKS~0637-752 quasar core dominates due to its soft spectrum,
the upper limits reach a minimum rather quickly, and generally
increase before becoming detections. It must be noted that the
detected Fermi emission in these bands is from the quasar core, not
the large-scale jet, based on the soft spectrum, and the fact that the
emission level rises as more bins are added (showing that the source
is indeed variable and that the bins are ordered by flux level). While
the 4th energy band detected points do not rise as quickly as the
first three, the flux level is far above the upper limit derived after
50 bins, so cannot be the steady emission of the large-scale jet,
which must be below this limit. The highest bin never shows a
significant detection of either component.

\subsection{\emph{Fermi} Analysis of 3C 273}

We re-analyzed the \emph{Fermi} data for 3C 273 using the 6-year
dataset to compare with the results from M14 using 4.5 years of data,
as the core remained relatively quiescient over the additional time
elapsed. We followed the same procedure as outlined above for
PKS~0637-752, finding six new sources within 7$^\circ$ of the position
of 3C~273, listed at right in Table~\ref{table:newsources}. The core of 3C~273
was detected with a TS of 17504, with a 100 MeV - 100 GeV photon flux
of 3.68$\times 10^{-7}$ s$^{-1}$ cm$^{-2}$ and $\Gamma_p$ = 2.67. A
lightcurve was made using bins totaling 10.5 days in GTI time, and
ordered according to TS (a total of 88 bins). The progressive-binning
was started from the single lowest bin, with the next-highest bin
continually added as described above until all bins were added. At
each step the flux (or 95\% upper limit) was calculated for the five
canonincal \emph{Fermi} energy bins.

\begin{table*}
\caption{\label{table:limits} Results of the Fermi Data Analysis}
\centering
\begin{tabular}{ccccccccccc}
\toprule
Source & Band & $E_1$ & $E_2$ & log Freq. & 95\% Limit              & Bins & Combined Knots\tablenotemark{*} & & Single Knot\tablenotemark{$\dagger$} & \\
       &      & (GeV) & (GeV) &   (Hz)    & (erg s$^{-1}$ cm$^{-2}$)  & Added &  $F_\mathrm{IC/CMB}$ & \% Ruled & $F_\mathrm{IC/CMB}$ & \% Ruled\\
       &      &       &       &           &                         & & (erg s$^{-1}$ cm$^{-2}$) & Out  & (erg s$^{-1}$ cm$^{-2}$) &  Out    \\
(1)    & (2)  & (3)   & (4)   & (5)       & (6) & (7) & (8) & (9) & (10) & (11) \\
\midrule
0637-752 & 1 & \phantom{00}0.1           & \phantom{00}{0.3}               & 22.6 & 1.05$\times 10^{-12}$     & 29     & 9.0$\times 10^{-13}$ & 92.9  & 3.6$\times 10^{-13}$ & ...  \\
             & 2 & \phantom{00}0.3           & \phantom{00}{1}\phantom{00} & 23.1 & 3.17$\times 10^{-13}$ & 32 & 8.8$\times 10^{-13}$ & 99.8  & 3.6$\times 10^{-13}$ & 94.5 \\
             & 3 & \phantom{00}1\phantom{00} & \phantom{00}{3}\phantom{00} & 23.6 & 1.80$\times 10^{-13}$ & 27 & 7.4$\times 10^{-13}$ & 99.98 & 3.2$\times 10^{-13}$ & 98.7 \\
             & 4 & \phantom{00}3\phantom{00} & \phantom{0}{10}\phantom{00} & 24.1 & 1.43$\times 10^{-13}$ & 50   & 5.3$\times 10^{-13}$ & 99.95 & 2.5$\times 10^{-13}$ & 98.6 \\
             & 5 & \phantom{0}10\phantom{00} &            100\phantom{00}  & 24.7 & 2.09$\times 10^{-13}$ & 75   & 2.3$\times 10^{-13}$ & 95.9  & 1.3$\times 10^{-13}$ & ...  \\
\\
3C 273       & 1 & \phantom{00}0.1           & \phantom{00}{0.3}           & 22.6 & 2.72$\times 10^{-11}$ & \phantom{0}1 & 2.1$\times 10^{-12}$ & ...      & 1.6$\times 10^{-12}$ & ...      \\
             & 2 & \phantom{00}0.3           & \phantom{00}{1}\phantom{00} & 23.1 & 4.63$\times 10^{-12}$ & \phantom{0}2 & 2.8$\times 10^{-12}$ & ...      & 2.1$\times 10^{-12}$ & ...      \\
             & 3 & \phantom{00}1\phantom{00} & \phantom{00}{3}\phantom{00} & 23.6 & 8.20$\times 10^{-13}$ & \phantom{0}5 & 3.6$\times 10^{-12}$ & $>$99.99 & 2.8$\times 10^{-12}$ & $>$99.99 \\
             & 4 & \phantom{00}3\phantom{00} & \phantom{0}{10}\phantom{00} & 24.1 & 4.46$\times 10^{-13}$ &          31  & 4.5$\times 10^{-12}$ & $>$99.99 & 3.5$\times 10^{-12}$ & $>$99.99 \\
             & 5 & \phantom{0}10\phantom{00} &            100\phantom{00}  & 24.7 & 3.56$\times 10^{-13}$ &          30  & 5.2$\times 10^{-12}$ & $>$99.99 & 4.1$\times 10^{-12}$ & $>$99.99 \\
\bottomrule
\end{tabular}
\end{table*}

\section{Results: Testing the IC/CMB Model}
\label{sec:results}

We show in Figure~\ref{fig:both_results} the radio to X-ray SEDs for
the jet of both PKS~0637-752 (left) and 3C~273 (right). For
PKS~0637-752 we have taken both the \emph{Chandra} X-ray and Hubble
Space Telescope (HST) infrared and optical data (NICMOS, WFPC2 and
ACS) from \cite{mehta2009}. We have also re-derived the \emph{Spitzer}
infrared fluxes for the brightest complex of knots (wk7.8, wk8.9,
wk9.7, wk10.7), following the same methods reported in
\cite{uchiyama2005}. 
We have also measured updated radio fluxes based on a re-analysis of
archival and new ATCA data at 4.8, 8.4, and 17.8 GHz. For 3C~273, data
is taken from \cite{jester2005,jes06} and \cite{uch06} and
references therein.

In both figures, we consider two scenarios: the first combines the
photometry of the brightest/nearest knots to the core in order to test
the IC/CMB prediction (black points and lines).  In 3C~273,
\cite{jester2005} have already shown that only knots A and B1 have
X-ray indices similar to their radio indices (which is required for
IC/CMB), so our ``combined knot'' scenario includes only these two
knots. For PKS~0637-752, we use all the bright X-ray knots just before
the turn in the jet (wk7.8, wk8.9, wk9.7, and wk10.6) where one might
assume that some deceleration likely takes place. The second scenario
assumes that the X-rays from the weaker knots are already \emph{not}
from IC/CMB, and so only the photometry of the X-ray brightest knot is
plotted (knot A in 3C~273 and wk8.9 in PKS~0637-752, plotted as blue
points and lines). The thin solid lines through radio-optical points
show a (phenomenological) synchrotron spectrum fitting the data, while
the heavy line shows the corresponding IC/CMB curve to match the X-ray
flux levels. As shown, for both jets, the 95\% upper limits in several
bands violate the IC/CMB predictions under either scenario.  



We report in the upper part of Table~\ref{table:limits} a summary of
the \emph{Fermi} data analysis for PKS~0637-752 and 3C~273. We list
the definition of the energy bins in columns 2-5, followed by the
deepest 95\% upper limit flux level (in $\nu F_\nu$) reached in our
progressive binning for each energy bin in column 6. The corresponding
number of bins co-added is given in column 7.  In column 8 we list the
flux predicted under the IC/CMB at the frequency given in column
5. This flux corresponds to the IC/CMB model prediction for the
combination of knots wk7.8, wk8.9, wk9.7, and wk10.7 in PKS~0637-752,
and knots A and B1 for 3C~273. In the former case, we do not include
the only other X-ray detected not (wk5.7) because it is not
consistently detected at other wavelengths. For 3C~273, only knots A
and B1 have X-ray spectra consistent with their radio spectra, so the
knots further downstream are assumed to be producing X-rays via
synchrotron emission. In column 10, we have calculated at what
significance level our observations rule out the level of predicted
IC/CMB flux given in column 9. For the final two columns, we also give
the predicted flux under IC/CMB and the significance-level that we can
rule it out, but only for the X-ray brightest knot of each jet (wk8.9
and knot A, respectively).  As shown, the IC/CMB model is ruled out at
a $>$ 99.9\% level for PKS~0637-752 and at $>$ 99.99\% level for
3C~273.

\section{Discussion}
\label{sec:discussion}
These two cases where the IC/CMB origin for the X-rays has been
unambiguously ruled out join with that of PKS~1136-135, where high UV
polarization has shown that the second component (UV to X-ray) must be
synchrotron in origin, since significant polarization is not expected
in the IC/CMB scenario \citep{cara2013}. we focus the rest of the
paper on the implications for jet physics if the X-ray flux in quasar
jets is synchrotron emission from a separate, high-energy electron
population.

\begin{figure*}[!t]
\begin{center}
\includegraphics[width=6.5in]{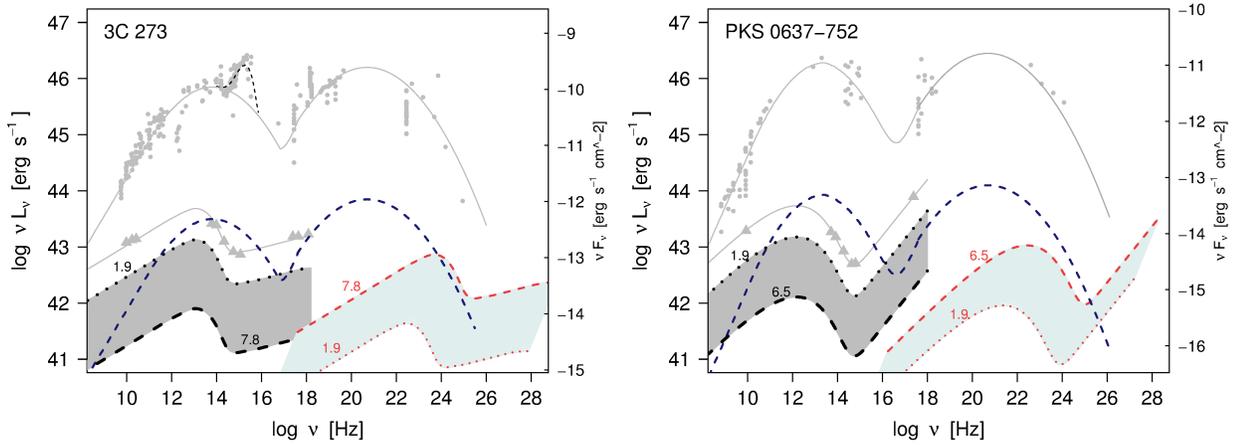}
\caption{\label{fig:tevplot}(Note: Figure description below applies to
  both panels; 3C~273 is shown at left and PKS~0637-752 is shown at
  right). The Core flux points are shown as gray circles with
  phenomenological SED fit through the points as a thin gray line (big
  blue bump shown as black dashed line and not included in the beamed
  emission fits). In comparison, the core points are shown as gray
  triangles.  The flux scale at right only applies to solid curves in
  the figure. The impression of the total dominance of the core flux
  is mainly a product of the beaming difference between the two
  components, as seen when beaming-corrected (angle-integrated)
  luminosities are plotted, rather than those assuming isotropy. Blue
  dashed line is the core fit times 1/$\delta^2$ with $\delta=15$ for
  both jets.  The gray zone is the range of possible beaming-corrected
  SEDs for the knots given our current constraints on $\delta$ for the
  knots.  Finally, the light blue shaded area is the range of IC/CMB
  emission possible given the same $\delta$ constraints. Note that the
  TeV emission in particular is already constrained to be much higher
  than is typical for `TeV blazars' ($\approx 10^{41}$ erg s$^{-1}$).}
\end{center}
\end{figure*}

A synchrotron origin for the X-rays is not in conflict with any of the
data in hand, and further, relaxes many of the `uncomfortable'
constraints of the IC/CMB model. Very small angles to the line of
sight are not required, and the total jet power required is
considerably less \citep{der04}, as the electron energy distribution
need not be extended to very low values. The main objection to a
second component heretofore has simply been its unexplained nature;
\cite{schwartz2000} notes that there is no reason why a second
population of high-energy electrons should be co-spatial with the
first. However, this co-occurance of two very different electron
populations, if the correct interpretation, is obviously a
very important clue to the particle acceleration mechanism in
large-scale jets, of which we still know little.

An interesting consequence follows for our accounting of the
large-scale-jet contribution to various backgrounds, especially at TeV
energies. Jet one-sidedness clearly indicates that the kpc-scale jets
are at least mildly relativistic, and thus IC/CMB emission must occur
at some level.  Due to the very low background in the highest-energy
\emph{Fermi} bands, the flux limits reachable by \emph{Fermi}'s
sky-scanning mode of operation should allow us to eventually either
detect this emission or put very strong limits on the factor of
$B/\delta$ which characterizes the flow on the kpc scale. The current
\emph{Fermi} upper limits already constrain $\delta\lesssim7.8$ for
3C~273 and $\delta\lesssim6.5$ PKS~0637-752, under the assumption of
equipartition magnetic fields, where we take $B\delta$ = 1.5$\times
10^{-5}$ G for PKS~0637-752 from \cite{tav00} and $B\delta$=1.0$\times
10^{-4}$ for 3C~273 from G06. These limits are already low enough to
have interesting consequences for our understanding of the total
radiative output of AGN jets on the kpc scale.

It is generally assumed that the radiative output of quasar jets is
dominated by that occurring at the `core', the base of the jet which
is presumed to be very near the black hole (or $\sim$ parsecs away at
most) and is therefore unresolved even in VLBI imaging. Certainly, the
observed fluxes are dominated by this part of the jet due to Doppler
beaming whenever the jet is pointed fairly along our
line-of-sight. This is depicted in Figure~\ref{fig:tevplot}, where the
core points of both jets are shown as gray circles, and the jet
photometry as gray triangles (flux scale on right axis). The
luminosity scale at left applies to these points only under the
incorrect assumption of isotropy.  If we correct these values for
beaming, we get the real `angle-integrated' total power output from
the core and the jet, plotted as a dark blue dashed line and the gray
shaded area, respectively. VLBI observations of superluminal motions
place a lower limit on $\Gamma$=15 for the cores of both 3C~273 and
PKS~0637-752, respectively \citep{lister2013,edwards2006}.  We have
applied a correction (multiplying by 1/$\delta^2$) assuming
$\delta$=15 for the cores of both jets to give the angle-integrated
core luminosity (dark blue dashed line). To calculate the
angle-integrated luminosity of the knots, we apply a lower limit value
of $\delta=1.9$ which comes from statistical arguments based on
populations \citep{arshakian2004}, which gives the dotted-line upper
edge to the gray shaded area, while the current $\delta$ limits from
\emph{Fermi} give the lower dashed-line limit.  The true
angle-integrated luminosity of the knots is thus somewhere in the gray
zone.

It is interesting to note that the knots are apparently \emph{not}
insignificant in total output when compared to the core. Definite
conclusions will require tighter constraints on the $\delta$ factors
of both the core and the knots, but it is possible that large-scale
jets contribute more than the core in the UV to X-rays, in addition to
their general dominance in the radio (which was already well
known). Large-scale jets could thus be an important contributor to
some astrophysical backgrounds.

A further observation follows from the realization that the X-rays are
synchrotron in origin: the electrons producing the synchrotron X-rays
will themselves upscatter the CMB to produce a GeV to TeV
spectrum. The angle-integrated total power in the IC/CMB component is
shown in Figure~\ref{fig:tevplot} as a light blue shaded area.  Note
that the bounds in this case are flipped; the upper $\delta$ limit
forms the upper edge of the allowed zone. Even in the minimum
$\delta$=1.9 case, these jets are already constrained to produce
fluxes in excess of $10^{41}$ erg s$^{-1}$ which is the typical total
radiative output for the canonical low-power FR I type `TeV
blazars'. We have not applied an EBL correction to these spectra
merely to illustrate the total intrinsic output. However, EBL
absorption is very important at TeV energies, and would make direct
observation of this TeV component very difficult. Even assuming the
most optimistic case of $\delta$=7.8 for 3C~273, at redshift 0.158 the
EBL absorption \citep{finke2010_ebl} is already high enough that it
would take at least 100 hours of observations by the future CTA
mission to detect the beamed IC/CMB component at TeV energies.  Thus
it is unlikely that many anomalous X-ray jets will have a synchrotron
origin for the X-rays directly confirmed via TeV observations, though
if \emph{Fermi} begins detecting the IC/CMB component, an upturn at
the highest energies might be visible in a few cases. The remaining
best direct observation is via polarization -- either in the UV, for
those that show the second component emerging there, or with future
X-ray polarimiters. Finally, we note that as long as \emph{Fermi}
continues to operate, the low background at the highest energies
should allow continually improving constraints on the $\delta$ factors
of large-scale jets.

\bigskip 
\begin{acknowledgments}
Work supported by Fermi Grant NNX13AO88G.
\end{acknowledgments}

\bigskip 
\bibliography{pks0637}

\end{document}